\documentstyle[12pt,aaspp4,flushrt,epsf]{article}

\def\be{\begin{equation}}
\def\ee{\end{equation}}
\def\beq{\begin{eqnarray}}
\def\eeq{\end{eqnarray}}
\def\half{{\textstyle{1\over2}}}
\def\ffrac#1#2{{\textstyle\frac{#1}{#2}}}
\def\sia{{\scshape sia}}
\def\sias{{\scshape sia}s}
\def\ria{{\scshape ria}}
\def\rias{{\scshape ria}s}
\def\bfp{{\bf p}}
\def\bfq{{\bf q}}
\def\bfP{{\bf P}}
\def\bfQ{{\bf Q}}

\def\bnabla{\mbox{\boldmath $\nabla$}}

\def\spose#1{\hbox to 0pt{#1\hss}}
\def\lta{\mathrel{\spose{\lower 3pt\hbox{$\sim$}}
    \raise 2.0pt\hbox{$<$}}}
\def\gta{\mathrel{\spose{\lower 3pt\hbox{$\sim$}}
    \raise 2.0pt\hbox{$>$}}}

\begin{document}

\title{Linear multistep methods for integrating reversible differential
equations} 

\author{N. Wyn Evans$^{1}$ and Scott Tremaine$^{2}$}

\medskip

\affil{$^1$Theoretical Physics, University of Oxford,  1 Keble
Road, \\ Oxford OX1 3NP, England}

\medskip

\affil{$^2$Princeton University Observatory, Peyton Hall, Princeton, 
NJ 08544, USA}

\begin{abstract}
This paper studies multistep methods for the integration of
reversible dynamical systems, with particular emphasis on the planar
Kepler problem. It has previously been shown by Cano \& Sanz-Serna
that reversible linear multisteps for first-order differential
equations are generally unstable.  Here, we report on a subset of
these methods -- the {\it zero-growth methods} -- that evade these
instabilities. We provide an algorithm for identifying these rare
methods. We find and study all zero-growth, reversible multisteps with six or
fewer steps.  This select group
includes two well-known second-order multisteps (the trapezoidal and
explicit midpoint methods), as well as three new fourth-order
multisteps -- one of which is explicit.  Variable timesteps can be readily
implemented without spoiling the reversibility.
Tests on Keplerian orbits show
that these new reversible multisteps work well on orbits with low or
moderate eccentricity, although at least 100 steps/radian are required
for stability. 
\end{abstract}

\section{Introduction}

The success of symplectic integration algorithms (\sias) as tools for
the numerical solution of Hamiltonian systems illustrates the
importance of numerical algorithms that inherit the fundamental
physical constraints of the systems to which they are applied
(``geometric integrators'').

Time-reversal symmetry plays a central role in physics (e.g., Davies
1974). To define reversibility formally and without reference to
coordinates, let $x$ denote the state of the system governed by the
differential equations
\be
{dx\over dt}=f(x).
\label{eq:basicde}
\ee
Let $T$ be an involution, i.e. an operator such that $T^2x=x$. The
system (\ref{eq:basicde}) is $T$-reversible if $T$ reverses the direction of
time, that is, if
\be
{dTx\over dt}=-f(Tx).
\ee
In standard phase space $x=(\bfq,\bfp)$ and $T=\hbox{diag}\,(1,-1)$. 

The trajectory of the system governed by the differential equations
(\ref{eq:basicde}) is defined by a map $G_t$ such that a system located at $x$ at
time 0 is found at $G_tx$ at time $t$; by definition 
$G_t=G_{-t}^{-1}$. Then, $T$-reversibility implies that
\be 
G_tTG_t=T \qquad \hbox{or} \qquad TG_t=G_t^{-1}T=G_{-t}T.  
\ee 

Reversible systems may or may not be Hamiltonian.
However, the general features of motion in reversible and Hamiltonian
systems show many similarities, including the existence of families of
KAM tori (e.g., Moser 1973; Arnold 1984; Roberts \& Quispel 1992).
These considerations motivate us to examine reversible integration
algorithms (\rias). A one-step numerical integration algorithm with
timestep $h$ is defined by an operator $\tilde{G}_h$ that is intended
to be a close approximation to $G_h$, and the algorithm is an \ria\ if
when it is applied to a reversible system 
\be
\tilde{G}_hT\tilde{G}_h=T.  
\ee 
Note that in contrast to $G_t$, $\tilde{G}_{h}$ is not necessarily
equal to $\tilde{G}_{-h}^{-1}$.

\subsection{Symplectic integration algorithms}

Let us first briefly review \sias; for more detail see Channell \& Scovel
(1990), Yoshida (1993), Marsden et al. (1996), and Sanz-Serna \& Calvo (1994).
The standard example of a Hamiltonian system is motion in a conservative
potential (this is also reversible). The Hamiltonian is
\be
H(\bfq,\bfp)=\half\bfp^2+U(\bfq)
\label{eq:sepham}
\ee
and the equations of motion are 
\be
{d\bfq\over dt}=\bfp, \qquad {d\bfp\over dt}=-\bnabla U,
\ee 
or 
\be
{d^2\bfq\over dt^2}=-\bnabla U.
\label{eq:second}
\ee
The most popular \sia\ for such systems is the leapfrog or Verlet
algorithm (e.g., Hockney \& Eastwood 1981). This is a map from the
phase-space coordinates $(\bfq_n,\bfp_n)$ at time $t_n$ to coordinates
$(\bfq_{n+1},\bfp_{n+1})$ at time $t_{n+1}=t_n+h$, defined by
\be 
\bfq'=\bfq_n+\half h\bfp_n, \quad
\bfp_{n+1}=\bfp_n-h\bnabla U(\bfq'), \quad
\bfq_{n+1}=\bfq'+\half h\bfp_{n+1}.  
\ee
Leapfrog is an explicit second-order method (i.e. the one-step error
is O$(h^3)$), but higher-order explicit \sias\ can be constructed by
concatenating leapfrog steps of different sizes, including backwards
steps. Leapfrog can be generalized to any separable Hamiltonian of the
form $H=\sum H_j$ where each $H_j$ is integrable.

For general Hamiltonians, \sias\ can be constructed by two
methods. The first is based on expanding the scalar generating
functions for the symplectic transformation $(\bfq_n,\bfp_n)
\to(\bfq_{n+1},\bfp_{n+1})$ in powers of $h$ (Kang 1986; Channell \&
Scovel 1990). The simplest example is the first-order \sia
\be
\bfq_{n+1}=\bfq_n+h{\partial H(\bfq_n,\bfp_{n+1})\over \partial \bfp_{n+1}},
\qquad \bfp_{n+1}=\bfp_n-h{\partial H(\bfq_n,\bfp_{n+1})\over \partial \bfq_n},
\ee
which is implicit in general but explicit for Hamiltonians of the form
(\ref{eq:sepham}). The second approach is to construct symplectic
implicit Runge-Kutta algorithms. For the differential equation
(\ref{eq:basicde}), the $s$-stage Runge-Kutta method is defined by
(e.g., Ralston \& Rabinowitz 1978)
\be
k_i=f(x_n+h\sum_{j=1}^sa_{ij}k_j),\qquad x_{n+1}=x_n+h\sum_{i=1}^s b_jk_j.
\label{eq:rkimp}
\ee
Runge-Kutta methods are symplectic if they satisfy a simple algebraic
test (Sanz-Serna \& Calvo 1994)
\be
b_ia_{ij}+b_ja_{ji}-b_ib_j=0, \qquad 1\le i,j \le s.
\label{eq:criterion}
\ee
On setting $i=j$ in (\ref{eq:criterion}), it is evident that all
symplectic Runge-Kutta methods are necessarily implicit. The best
known examples are the Gauss-Legendre Runge-Kutta methods (e.g.,
Sanz-Serna \& Calvo 1994). The simplest can be written
\be
x_{n+1}=x_n+hf\left[\half(x_n+x_{n+1})\right],
\label{eq:midptt}
\ee 
which is the second-order implicit midpoint method. A symplectic
fourth-order two-stage Runge-Kutta method is given by
\be
a_{11}=a_{22}=\ffrac{1}{4}, \quad 
a_{12}=\ffrac{1}{4}-\ffrac{1}{6}\sqrt{3}, \quad 
a_{21}=\ffrac{1}{4}+\ffrac{1}{6}\sqrt{3}, \qquad 
b_1=b_2=\ffrac{1}{2}.
\label{eq:rk}
\ee
More generally, the Gauss-Legendre Runge-Kutta method is the unique
$s$-stage method with order $2s$, and this method is always
symplectic.  These attractive features are offset by the computational
cost of solving the implicit equations (\ref{eq:rkimp}).

A limitation of \sias\ is that they are difficult to generalize to
variable timesteps. When a standard variable timestep prescription is
applied to an \sia, its performance is no better than that of
non-symplectic integrators. The reason is simply that the mapping
$G_{h(x)}(x)$ is usually not symplectic even when $G_h(x)$ is.  This
is a serious problem, since most applications benefit from a variable
timestep. One way to introduce variable timestep is by
extending the phase space (Mikkola 1997). Suppose we wish to use a
timestep $g(\bfq,\bfp)$. We define an extended phase space
$(\bfQ,\bfP)=((q_0,\bfq),(p_0,\bfp))$ by $q_0=t$ and $p_0=-E$, where
$t$ is time and $E$ is energy. We then define a new Hamiltonian
\be
\Gamma(\bfQ,\bfP)=g(\bfq,\bfp)\left[H(\bfq,\bfp)+p_0\right].
\ee
The equations of motion for the Hamiltonian $\Gamma$ in the extended
phase space with fictitious time $\tau$ are the same as the equations
of motion for the Hamiltonian $H$ in the original phase space,
supplemented by the condition $dt/d\tau=g(\bfq,\bfp)$. We may now
integrate the Hamiltonian $\Gamma$ using an \sia\ with fixed timestep
$\Delta\tau=1$, which corresponds to $\Delta t\simeq g$. A limitation
of this approach is that the Hamiltonian $\Gamma$ is generally not
separable, so that leapfrog and its generalizations cannot be applied.

\subsection{Reversible integration algorithms}

Several well-known one-step second-order algorithms for the
differential equation (\ref{eq:basicde}) are \rias: for example, the
implicit midpoint method (eq. \ref{eq:midptt}), the trapezoidal
method,
\be
x_{n+1}=x_n+\half h\left[f(x_n)+f(x_{n+1})\right],
\label{eq:trapp}
\ee
and the explicit midpoint method,
\be
x_{n+2}=x_n+hf\left(x_{n+1}\right).
\label{eq:expmidpt}
\ee
Some \sias\ are also \rias\ when they are applied to reversible
Hamiltonians. Leapfrog and its generalizations are reversible.  A
reversible version of leapfrog with variable timestep $g(\bfq,\bfp)$
is given by (Huang \& Leimkuhler 1997; Calvo et al. 1998)
\begin{eqnarray}
\bfq' = \bfq_n+{h\over 2\rho_n}\bfp_n, &\qquad&
\bfp' = \bfp_n+{h\over 2\rho_n}\bnabla U(\bfq'), \nonumber \\
\rho_{n+1} = {2\over g(\bfq',\bfp')}-\rho_n, &\qquad&
\bfp_{n+1} = \bfp'+{h\over 2\rho_{n+1}}\bnabla U(\bfq'), \nonumber \\
\bfq_{n+1} = \bfq'+{h\over 2\rho_{n+1}}\bfp_{n+1}, &\qquad&
t_{n+1} = t_n+{h\over 2}\left(\rho_n^{-1}+\rho_{n+1}^{-1}\right).
\label{eq:calvo} 
\end{eqnarray} 
Quinn et al. (1997) describe a closely related algorithm based on a
discrete set of timesteps that depend only on $\bfp$ or $\bfq$ and are
separated by factors of two.  

A different approach to constructing \rias\ is to modify non-reversible
integration algorithms. For example, the following two operators are
reversible even if $\tilde{G}_h$ is not:
\be
\tilde{G}_{h/2}T\tilde{G}_{h/2}^{-1}T, \qquad
(1+T\tilde{G}_hT)^{-1}(1+\tilde{G}_h);
\ee
if we write $\tilde{G}_h=1+hA+\hbox{O}(h^2)$ then both of these
operators are equal to $1+\half h(A-TAT)$ to first order in $h$. Hut
et al. (1997) describe numerical experiments with the second of these
operators, calling it a ``time-symmetrization meta-algorithm'' since
it can be applied to any one-step numerical integration algorithm.
Reversible algorithms remain reversible with a variable timestep so
long as the timestep is determined symmetrically by the location of
the system at the start and end of the step, e.g. $h=\half[g(x_n)+
g(x_{n+1})]$ (Hut et al. 1995).

Special second-order differential equations of the form
\be 
{d^2x\over dt^2}=F(x)
\label{eq:fff}
\ee 
are reversible, and so can profitably be integrated using \rias. A useful
source of high-order \rias\ is linear multistep methods (e.g., Henrici 1962;
Gear 1971; Lambert 1973), which have the form
\be
\sum_{j=0}^k \left(a_{k-j}x_{n-j}-h^2b_{k-j}F_j\right)=0,
\ee 
where $x_j=x(t_0+jh)$, $F_j=F(x_j)$. We may assume without loss of
generality that $a_k=1$. The method is explicit if $b_k=0$ and
otherwise implicit. Linear multisteps include the classic St\"ormer
(explicit) and Cowell (implicit) methods, which are characterized by
$a_k=1$, $a_{k-1}=-2$, $a_{k-2}=1$, and $a_{k-j}=0$ for $j>2$.  It is
easy to show that the requirement for reversibility is that
$a_j=ca_{k-j}$ and $b_j=cb_{k-j}$ where $c=\pm 1$; thus the St\"ormer
and Cowell methods are generally not reversible.  Multistep \rias\ are
discussed by Lambert \& Watson (1976), Quinlan \& Tremaine (1990), and
Fukushima (1998, 1999). In general, their performance over long time
intervals on reversible dynamical systems is much better than
St\"ormer-Cowell methods, although for occasional unfortunate choices
of timestep their performance is ruined by timestep resonances
(Quinlan 1999).

Systems of first-order differential equations such as (\ref{eq:basicde})
can be integrated by linear multistep methods of the form
\be
\sum_{j=0}^k\left(\alpha_{k-j}x_{n-j}-h\beta_{k-j}f_{n-j}\right)=0,
\label{eq:basic}
\ee 
where $x_j=x(t_0+jh)$, $f_j=f(x_j)$. We can assume without loss of
generality that $\alpha_k=1$; explicit methods have $\beta_k=0$. These
include the classic Adams-Bashforth (explicit) and Adams-Moulton
(implicit) methods, which have $\alpha_{k-1}=-1$ and $\alpha_{k-j}=0$
for $j>2$ (e.g., Henrici 1962; Gear 1971).  Multistep methods for
first-order differential equations are more general than multistep
methods for special second-order equations, since any second-order
equation can be written as a set of first-order equations. Moreover,
implementing variable timesteps is easy in first-order equations -- if
we wish to use a timestep $g(x)$ we introduce a fictitious time $\tau$
by the relation $dt=g(x)d\tau$, and equation (\ref{eq:basicde})
can be rewritten as
\be
{dx\over d\tau}=g(x)f(x),
\ee
which can be integrated using unit timestep in $\tau$. 

The aim of this paper is to investigate linear multistep \rias\ for
first-order differential equations. An important earlier investigation is that
of Cano \& Sanz-Serna (1998), who found that such methods typically possess
grave numerical instabilities.  We review general linear multistep methods in
\S \ref{sec:intro}, multistep \rias\ in \S \ref{sec:reverse}, together with
their instabilities in \S \ref{sec:zero}. We show in \S \ref{sec:zmm} that it
is possible to construct some linear multistep \rias\ that are not subject to
the Cano \& Sanz-Serna instabilities. A general discussion of stable multistep
\rias\ with up to 6 steps is given in \S \ref{sec:onetwo} -- \S
\ref{sec:fivesix}. Finally, \S \ref{sec:kep} describes numerical examples
based on integrating Kepler orbits and \S \ref{sec:disc} discusses our
results.

\section{Linear multistep methods}

\label{sec:intro}

Following Henrici (1962) and Lambert (1973), we associate with
(\ref{eq:basic}) the linear operator
\be
L[x(t),h]=\sum_{j=0}^k\big\{\alpha_{k-j}x[t+(k-j)h]-
h\beta_{k-j}x'[t+(k-j)h]\big\}.
\label{eq:linearoperator}
\ee
We can expand $x(t)$ in a Taylor series to obtain
\be
L[x(t),h]=\sum_{q=0}^\infty C_qh^qx^{(q)}(t),
\label{eq:tay}
\ee
where
\be
C_0=\sum_{l=0}^k\alpha_l,\qquad C_q={1\over q!}\sum_{l=0}^k\alpha_ll^q - 
{1\over(q-1)!}\sum_{l=0}^k\beta_ll^{q-1},\quad q=1,2,\ldots
\ee
The order $p$ of the multistep is the unique integer for which
$C_0=\cdots=C_p=0$, $C_{p+1}\not=0$. The constant
$C=C_{p+1}/\sigma(1)$ is known as the error constant and is a measure
of the local truncation error. Now define the characteristic
polynomials 
\be
\rho(\xi)=\sum_{j=0}^k\alpha_j\xi^j,\qquad\sigma(\xi)=
\sum_{j=0}^k\beta_j\xi^j.
\label{eq:poly}
\ee
If $x(t)=\exp(\lambda t)$, then
\be
L[x(t),h]=e^{\lambda t}[\rho(e^{\lambda h})-\lambda h\sigma(e^{\lambda
h})],
\ee
which together with equation (\ref{eq:tay}) implies
\be
\rho(1+z)-\log(1+z)\sigma(1+z)=C_{p+1}z^{p+1}+\hbox{O}(z^{p+2}),
\label{eq:series}
\ee
where $p$ is the order and $z=\exp(\lambda h)-1$. Order $\ge 0$
requires
\be
\rho(1)=\sum_{j=0}^k\alpha_j=0,
\label{eq:rhoone}
\ee
and order $\ge 1$ requires
\be
\rho'(1)=\sigma(1).
\label{eq:rel}
\ee
A multistep method is zero-stable if and only if the roots of
$\rho(\xi)$ lie inside the unit circle in the complex plane, or are on
the unit circle and simple (proofs of this are given by Henrici 1962,
Gear 1971 and Lambert 1973). Zero-stability ensures that the parasitic
solutions generated by the additional roots of the difference equation
(which is of order $k$, while the original differential equation
(\ref{eq:basicde}) has order one) do not grow, at least in the
limit of zero timestep. We will discuss other forms of stability
shortly in \S \ref{sec:zero}.

We assume that the polynomials $\rho(\xi)$ and $\sigma(\xi)$ have no
common roots other than 1, for the following reason. Suppose that
$\xi_0\not=1$ is a common root, so that $(\xi-\xi_0)$ is a common
factor. Then $\rho(\xi)=(\xi-\xi_0)\tilde\rho(\xi)$,
$\sigma(\xi)=(\xi-\xi_0)\tilde\sigma(\xi)$ where $\tilde\rho(\xi)$ and
$\tilde\sigma(\xi)$ are polynomials of degree $k-1$. Then equation
(\ref{eq:series}) may be written
\be
\tilde\rho(1+z)-\log(1+z)\tilde\sigma(1+z)={C_{p+1}\over 1-\xi_0}z^{p+1}+
\hbox{O}(z^{p+2}).
\label{eq:seriesa}
\ee
Thus $\tilde\rho(\xi), \tilde\sigma(\xi)$ define a $(k-1)$-step method
with the same order and the same error constant
($C=C_{p+1}/\sigma(1)=C_{p+1}/[(1-\xi_0)\tilde\sigma(1)]$) as the
original $k$-step method, and there is no obvious reason why the
simpler method should not be used instead.

There are $2k+1$ unknown coefficients in equation (\ref{eq:basic})
(since $\alpha_k=1$), and therefore in principle these coefficients
can be chosen so that the order is $2k$ (or $2k-1$ if the method is
explicit, with $\beta_k=0$). However, the maximum order of a
zero-stable multistep method is $k+1$ if $k$ is odd and $k+2$ if $k$
is even (Henrici 1962).

\subsection{Reversible multistep methods}

\label{sec:reverse}

Suppose that a trajectory $\{x_{n-k},x_{n-k+1},\ldots,x_n\}$,
$\{f_{n-k},f_{n-k+1},\ldots,f_n\}$ satisfies the difference equation
(\ref{eq:basic}). The same segment of the time-reversed trajectory is
given by $\{Tx_n,Tx_{n-1},\ldots,Tx_{n-k}\}$,
$\{g_n,g_{n-1},\ldots,g_{n-k}\}$, where $T$ is the time-reversal
operator and $g_k=f(Tx_k)$. We shall assume that $x_k=({\bf r}_k, {\bf
v}_k)$ and that $f_k=({\bf v}_k,-\bnabla\Phi({\bf r}_k))$. Then
$T=\hbox{diag}\,(1,-1)$ and $f(Tx)=-Tf(x)$.  In an \ria\ the time-reversed
trajectory should also satisfy the difference equation
(\ref{eq:basic}), that is
\be
\sum_{j=0}^k(\alpha_{k-j}Tx_{n-k+j}-h\beta_{k-j}g_{n-k+j})=0.
\ee
Since $g_j=f(Tx_j)=-Tf_j$, we may operate with $T$ to obtain
\be
\sum_{j=0}^k(\alpha_{k-j}x_{n-k+j}+h\beta_{k-j}f_{n-k+j})=0,
\ee
or, equivalently,
\be
\sum_{j=0}^k(\alpha_jx_{n-j}+h\beta_jf_{n-j})=0.
\label{eq:reverse}
\ee
If (\ref{eq:reverse}) is to be satisfied whenever (\ref{eq:basic}) is
satisfied, then we must have
\be
\alpha_{k-j}=c\alpha_j,\qquad \beta_{k-j}=-c\beta_j,
\label{eq:cond}
\ee
where $c$ is a constant. Applying this relation twice gives $c^2=1$ so
$c=\pm 1$. If $c=+1$, we shall say that the multistep method has even
parity; if $c=-1$, the parity is odd.  The characteristic polynomials
(\ref{eq:poly}) now satisfy
\be
\rho(\xi)=c\xi^k\rho(\xi^{-1}),\qquad 
\sigma(\xi)=-c\xi^k\sigma(\xi^{-1})
\label{eq:symm}.
\ee
Now let $\xi_j, j=1,\ldots,k$ be the roots of $\rho(\xi)$. Equation
(\ref{eq:symm}) implies that if $\xi_j$ is a root then so is
$\xi_j^{-1}$. For zero-stability, the roots cannot lie outside the
unit circle -- therefore they must lie on the unit circle, and
moreover they must be simple.  Equation (\ref{eq:rhoone}) guarantees
that $\xi_1=1$ is a root for any method with order $\ge 0$. This root
is simple if $\rho'(1)\not =0$, and equation (\ref{eq:rel}) then
requires that $\sigma(1)\not=0$ for any method with order $\ge
1$. However, even-parity methods have $\sigma(1)=\sum_{j=0}^k\beta_j=0$.
Therefore, even-parity methods are not zero-stable and we can restrict
our attention to odd-parity methods, $c=-1$.

If the reversibility criterion (\ref{eq:cond}) is satisfied, then
\be
L[x(t),h]=cL[x(t+kh),-h].
\ee
Using (\ref{eq:tay}) and the Taylor series expansion for $x(t+kh)$, we
may derive the constraint
\be
C_p=c\sum_{n=0}^p{(-1)^{p-n}k^n\over n!}C_{p-n}.
\ee
We can re-write this equation as
\be
C_p[1-c(-1)^p]=\sum_{n=1}^p{(-1)^{p-n}k^n\over n!}C_{p-n}.
\ee
If $c(-1)^p=-1$, then this equation implies that $C_p=0$ if
$C_0=\cdots=C_{p-1}=0$ and thus the order cannot be $p-1$. Therefore,
odd-parity methods have even order. 

The stability of multistep methods for oscillatory problems can be
parametrized by the interval of periodicity introduced by Lambert \&
Watson (1976). Suppose the right-hand size of equation
(\ref{eq:basicde}) is $f(x,t)=i\omega x$, with $\omega$ real. Then,
the multistep method (\ref{eq:basic}) becomes a linear difference
equation, with solution $x_n=a\xi^n$. Here, $a$ and $\xi$ are complex
constants, the latter satisfying
\be
\rho(\xi)-i\omega h\sigma(\xi)=0\qquad \hbox{or}\qquad g(\theta)\equiv 
-i{\rho(e^{i\theta})\over\sigma(e^{i\theta})}=\omega h.
\label{eq:period}
\ee
The interval of periodicity is the largest value of $\omega h$ such
that all of the roots of the first of equations (\ref{eq:period}) lie
on the unit circle. Outside the interval of periodicity, the solution
grows exponentially and hence is unstable.  For reversible methods
with odd parity, we may write
\be
g(\theta)={\sum_{j=0}^k\alpha_j\sin[(j-\half k)\theta]\over
\sum_{j=0}^k\beta_j\cos[(j-\half k)\theta]}.
\ee
If the multistep method is zero-stable, then $\rho(\xi)$ has $k$
distinct roots on the unit circle. Thus, $g(\theta)$ has $k$ distinct
roots on the interval $[-\pi,\pi]$.  For sufficiently small values of
$\omega h$, the equation $g(\theta)=\omega h$ will still have $k$
roots. As $\omega h$ is increased, eventually a pair of these roots
will disappear. This occurs at the smallest local maximum of
$g(\theta)$ and marks the end of the interval of periodicity. So, a
plot of $g(\theta)$ can be used to determine the interval of
periodicity (Fukushima 1998).

\subsection{Instabilities in reversible multisteps}

\label{sec:zero}

The growth of errors in multistep methods is discussed by Henrici
(1962), Gear (1971) and especially by Cano \& Sanz-Serna (1998). For
our purposes, a limited heuristic version of these analyses is sufficient.  Let
$\{x_j\}$ be a solution of the multistep method (\ref{eq:basic}). Now
perturb the solution to $\{x_j+e_j\}$, $|e_j|\ll |x_j|$.  Linearizing
equation (\ref{eq:basic}), we have
\be
\sum_{j=0}^k(\alpha_{k-j}-h\beta_{k-j}f'_{n-j})e_{n-j}=0,
\label{eq:basica}
\ee
where $f'_j=\partial f(x_j)/\partial x$. Now look for solutions of the
form $e_n=\xi^ny(t_0+nh)$ where $y(t)$ is smooth and $\xi$ is a complex
constant. In the limit $h\to 0$, $x_n\to \tilde x(t_0+nh)$, where $\tilde
x(t)$ is the accurate solution of the differential equation (\ref{eq:basicde})
and therefore is smooth. Then (\ref{eq:basica}) becomes
\be
\sum_{j=0}^k\left[\alpha_j-h\beta_j{\partial f\over\partial x}
(\tilde x(t+jh))\right]\xi^jy(t+jh)=0,
\label{eq:basicb}
\ee
where $t=t_0+(n-k)h$. Now expand in a Taylor series in $h$. To order 
$h^0$ we have $y(t)\rho(\xi)=0$, which requires that $\xi$ is one of the roots
$\xi_j, j=1,\ldots,k$ of the polynomial $\rho$. To order $h$ we have
\be
{dy(t)\over dt}=\left(\sum_{j=0}^k\beta_j\xi^j\over 
\sum_{j=0}^kj\alpha_j\xi^j\right)y(t)
{\partial f\over\partial x}(\tilde x(t))=
\left[\sigma(\xi)\over\xi\rho'(\xi)\right]
{\partial f\over\partial x}(\tilde x(t)).
\ee
Thus, as $h\to0$, the linearized difference equation (\ref{eq:basica})
has solutions of the form
\be
e_n=\sum_{j=1}^k a_j\xi_j^ny_j(t_0+nh),
\ee
where $y_j(t)$ satisfies the differential equation
\be
{dy_j\over dt}=\lambda_j{\partial f(\tilde x(t))\over\partial x}y_j(t),
\label{eq:lvar}
\ee
and the growth parameters are 
\be
\lambda_j={\sigma(\xi_j)\over\xi_j\rho'(\xi_j)},\qquad j=1,\ldots,k.
\label{eq:growth}
\ee
This should be contrasted with the variational equation of the orbit
itself. If the trajectory $\tilde x(t)$ is perturbed to $\tilde
x(t)+\tilde e(t)$, $|\tilde e|\ll |\tilde x|$, then
\be
{d\tilde e(t)\over dt}={\partial f(\tilde x(t))\over\partial x}\tilde e(t).
\label{eq:var}
\ee
Cano \& Sanz-Serna (1998) point out that the variational equations
(\ref{eq:lvar}) and (\ref{eq:var}) generally have quite different
solutions, and hence the former can be unstable even if the latter is
stable.  However, this instability can be evaded for special
values of the growth parameters $\lambda_j$: for $\lambda_j=1$,
equations (\ref{eq:growth}) and (\ref{eq:var}) are the same, for
$\lambda_j=-1$, equation (\ref{eq:growth}) is simply the time-reversed
version of equation (\ref{eq:var}), and for $\lambda_j=0$, the
solution of equation (\ref{eq:growth}) is simply $y_j=$const. We use
the term ``zero-growth'' methods to denote multistep methods such that
$\lambda_j\in \{-1,0,1\}, j=1,\ldots,k$. Cano (1996) has shown that
for the reversible planar Kepler problem, these are the only possible
choices for the growth parameters $\lambda_j$ to ensure stability,
although for other potentials this may not be the case.

Multistep methods for integrating the special second-order equation
$x''=f(x,t)$ (Lambert \& Watson 1976; Quinlan \& Tremaine 1990) are not
generally subject to this sort of instability, for the following reason. The
expansion of the analog to equation (\ref{eq:basicb}) in this case yields
$y(t)\rho(\xi)=0$ to order $h^0$ and $y'(t)\rho'(\xi)=0$ to order
$h$. Satisfying these two equations simultaneously requires that $\xi$ is a
double root of $\rho(\xi)$; therefore there is no instability if all of the
roots of $\rho(\xi)$ are simple (Cano \& Sanz-Serna 1998).

\section{Zero-growth methods}

\label{sec:zmm}

In this section, all the zero-growth multistep algorithms with up to six
steps are found. Let us begin with some general results for 
stable, odd-parity, multistep \rias, specified by the characteristic
polynomials $\rho(\xi)$ and $\sigma(\xi)$. Let $\xi_j, j=1,\ldots,k$
be the roots of $\rho(\xi)$; for zero-stable reversible methods, these
are distinct, lie on the unit circle and are either real ($+1$ or
$-1$) or appear in complex-conjugate pairs. Equation (\ref{eq:rhoone})
implies that $1$ is always a root, which we denote $\xi_1$.
Therefore, $-1$ is also a root if and only if the total number of
roots $k$ is even; when present, we denote this root as $\xi_2$.  We
normalize the multistep method so that $\alpha_k=1$ and
\be
\rho(\xi)=\prod_{j=1}^k(\xi-\xi_j).  
\ee
We now demand that the growth parameter associated with the root
$\xi_l$ is
$\lambda_l\in\{-1,0,1\}$ (note that $\lambda_1=+1$ by
eq. \ref{eq:rel}). From equation (\ref{eq:growth}), this requires that
\be
\sigma(\xi_l)=\lambda_l\xi_l\rho'(\xi_l)=\lambda_l\xi_l\prod_{j\not=l}(\xi_l-\xi_j).
\label{eq:sigjj}
\ee
The unique polynomial of order $k$ that passes through the points
(\ref{eq:sigjj}) and satisfies the symmetry condition (\ref{eq:symm}) is
\be
\sigma(\xi)=\half \sum_{l=1}^k \lambda_l(\xi+\xi_l)\prod_{j\not=l}(\xi-\xi_j),
\label{eq:sigdef}
\ee
with $\lambda_j=\lambda_l$ when $\xi_j\xi_l=1$. An alternative form is
\be
\sigma(\xi)=\half\rho(\xi)\sum_{l=1}^k \lambda_l{\xi+\xi_l\over \xi-\xi_l}.
\label{eq:sigdefa}
\ee

Note that if $\lambda_l=0$ then $\xi_l$ is a root of both $\rho(\xi)$
and $\sigma(\xi)$. In this case, there is a zero-growth multistep
\ria\ with the same order and error constant but fewer steps (see \S
\ref{sec:intro}). Therefore, we restrict our attention to methods with
$\lambda_l\in\{-1,1\}$.

The method is explicit if $\sigma(0)=0$, which in turn requires
\be
\sum_{j=1}^k \lambda_j=0,
\label{eq:www}
\ee 
a condition that can only be satisfied if $k$ is even. 

A $k$-step method is completely specified by the $k^{\rm th}$-order
polynomials $\rho(\xi)$, $\sigma(\xi)$. In a stable multistep \ria\
with $k$ even, the roots of $\rho(\xi)$ are $+1$, $-1$, $x_j\pm iy_j$
where $x_j^2+y_j^2=1$, $j=1,\ldots \half(k-2)$. For $k$ odd, the root
at $-1$ is not present and the index runs to $\half(k-1)$. In a
zero-growth method, the polynomial $\sigma(\xi)$ is determined by
equation (\ref{eq:sigdef}) once $\rho(\xi)$ and the signs $\lambda_j$ are
specified.  We now discuss all stable, zero-growth, multistep \rias\
of $\le 6$ steps. We label the interesting methods by SZ$k$p where $k$
is the number of steps and ``p'' is an optional suffix that
distinguishes different methods with the same $k$.

\begin{figure}
\vspace{12.5cm}
\includegraphics{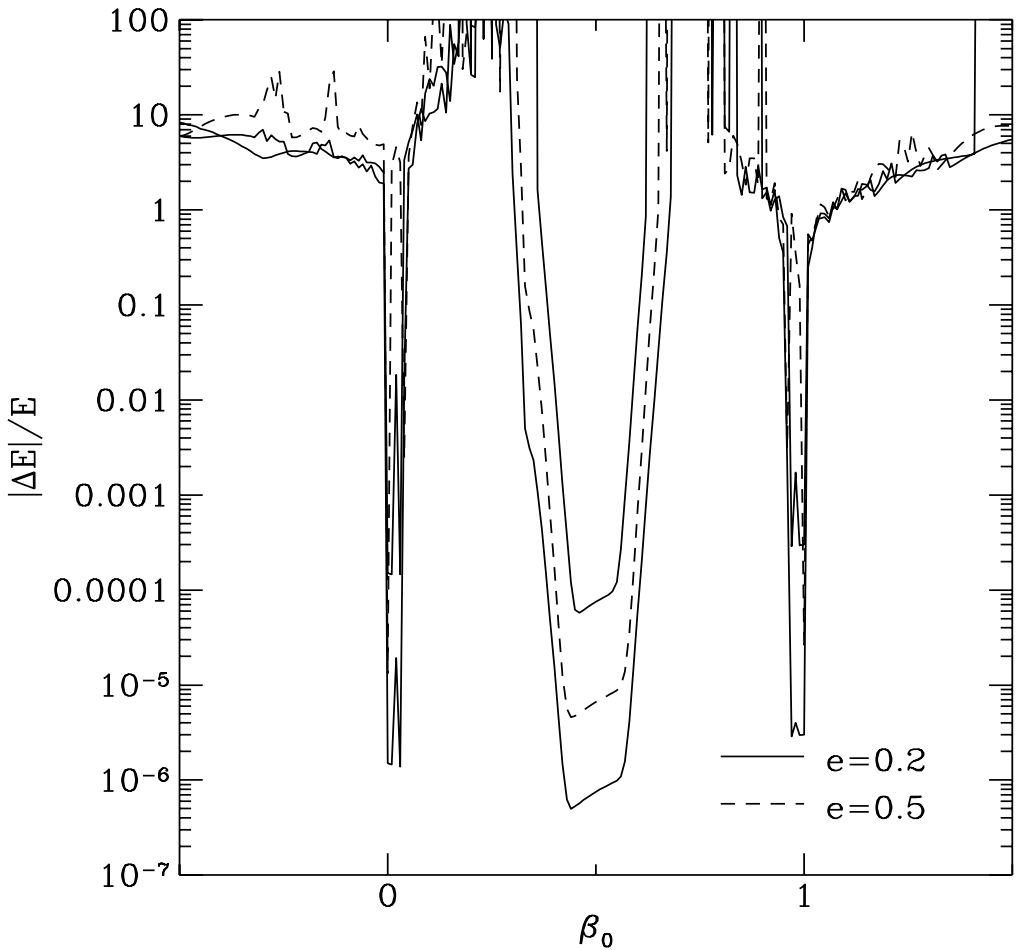}
\caption{The fractional energy error after integrating the standard
Kepler problem ($a=1$, $GM=1$) for 100 time units using the 2-step
\ria\ (\ref{eq:btwo}), as a function of the parameter $\beta_0$. The
two solid curves were computed with timesteps $h=0.01$ and $0.001$
for eccentricity $e=0.2$, while the dashed curve has $h=0.001$ and
$e=0.5$. Off-scale energy errors indicate that an implicit timestep
did not converge after 20 iterations. The minima at
$\beta_0=0,\half,1$ correspond to the explicit midpoint method, the
trapezoidal method, and the trapezoidal method with a timestep of
$2h$.}
\label{fig:multi2}
\end{figure}

\subsection{One- and two-step methods}

\label{sec:onetwo}

Since $\xi_1=1$ is always a root, for $k=1$ we must have
$\rho(\xi)=\xi-1$. Since $\lambda_1=+1$ by equation (\ref{eq:rel}), the
zero-growth requirement is automatically satisfied. Equation
(\ref{eq:sigdef}) requires that $\sigma(\xi)=\half(\xi+1)$, which
yields the trapezoidal method (eq. \ref{eq:trapp}) 
\be
x_{n+1}=x_n+\half h(f_{n+1}+f_n); \qquad\qquad\hbox{method SZ1}
\label{eq:trap}
\ee
The first few terms of the Taylor series (\ref{eq:tay}) are
\be
C_0=0, \quad C_1=0, \quad C_2=0, \quad C_3=-\ffrac{1}{12}; 
\ee
thus the error constant is $C=-\frac{1}{12}$ and the method is second-order.
The function $g(\theta)$ (eq. \ref{eq:period}) is $2\tan\half\theta$, so the
interval of periodicity is infinite.

For $k=2$ the roots of $\rho(\xi)$ can only be $\pm 1$ ($\xi_1=1$ is
always a root; $\xi_2$ must be real since $\xi_2^*$ is also a root and
there are only two roots; $\xi_2$ must be on the unit circle and
distinct from $\xi_1$; thus $\xi_2=-1$). Thus, the first
characteristic polynomial is
\be
\rho(\xi)=\xi^2-1.
\ee
There are two choices for the second characteristic polynomial
$\sigma(\xi)$, depending on $\lambda_2\in \{-1,1\}$:
\be
\sigma(\xi)=2\xi,\qquad\hbox{or}\qquad  \sigma(\xi)=\xi^2+1.
\label{eq:sigtwo} 
\ee 
The first of these gives the explicit midpoint method (eq. \ref{eq:expmidpt}),
\be
x_{n+1}=x_{n-1}+2hf_n \qquad\qquad\hbox{method SZ2}.
\label{eq:mid} 
\ee
The first few terms of the Taylor series (\ref{eq:tay}) are
\be
C_0=0, \quad C_1=0, \quad C_2=0, \quad C_3=\ffrac{1}{3}; 
\ee
thus the error constant is $C=\frac{1}{6}$ and the method is second-order.
The function $g(\theta)$ is $\sin\theta$ so the interval
of periodicity is $(\omega h)_{\rm max}=1$.  The second equation for
$\sigma(\xi)$ in (\ref{eq:sigtwo}) yields
\be 
x_{n+1}=x_{n-1}+h(f_{n+1}+f_{n-1}), 
\label{eq:traptwo}
\ee 
which is the same as the trapezoidal method (\ref{eq:trap}), with a
timestep of $2h$.

All of these two-step methods are special cases of a one-parameter
family of multistep \rias,
\be
x_{n+1}=x_{n-1}+h[\beta_0f_{n+1}+2(1-\beta_0)f_n+\beta_0f_{n-1}].
\label{eq:btwo}
\ee
The explicit midpoint method (eq. \ref{eq:mid}) has $\beta_0=0$ and
the trapezoidal method with timestep $2h$ (eq. \ref{eq:traptwo}) has
$\beta_0=1$.  When $\beta_0=\frac{1}{2}$, the method is simply the
composition of two trapezoidal steps (eq. \ref{eq:trap}).  Milne's
method ($\beta_0=\frac{1}{3}$) is the two-step method of maximum
possible order ($p=4$), but its disadvantage is that its growth
parameter $\lambda_2=-\frac{1}{3}$ and so it is susceptible to
numerical instability.  The interval of periodicity $(\omega h)_{\rm
max}=(1-2\beta_0)^{-1/2}$ for $\beta_0<\half$ and infinity for
$\beta_0\ge\half$.  To compare these methods we have integrated a
Kepler orbit with unit semimajor axis and period $2\pi$ 
for 100 time units using various
eccentricities and timesteps (Fig. \ref{fig:multi2}). Even for
timesteps as low as $h=0.001$, the relative energy error $|\Delta
E|/E\gg 1$, except near the zero-growth methods at
$\beta_0=0,\half,1$. In particular, the second-order zero-growth
methods -- even the explicit method at $\beta_0=0$ -- exhibit much
better behavior than the fourth-order Milne's method.

\subsection{Three- and four-step methods}

\label{sec:threefour}

For $k=3$, the distinct roots of $\xi$ can only be $\xi_1=1$ and
$\xi_{2,3}=u\pm iv$ where $|u|<1$ and $v=(1-u^2)^{1/2}$. Thus,
\be
\rho(\xi)=\xi^3-(2u+1)\xi^2+(2u+1)\xi-1.  
\ee 
There are two choices for $\sigma(\xi)$. The first choice is to take
$\lambda_2=\lambda_3=+1$, so that we have
\be
\sigma(\xi)=\ffrac{3}{2}\xi^3-(\half+u)\xi^2-(\half+u)\xi+\ffrac{3}{2}.
\ee 
The first few terms of the Taylor series (\ref{eq:tay}) are 
\be 
C_0=0, \quad C_1=0, \quad C_2=0, \quad C_3=-\ffrac{13}{6}+\ffrac{1}{6}u. 
\ee 
$C_3=0$ has no solution for $|u|<1$ so the method is second-order; the
interval of periodicity is infinite. The second choice is to take
$\lambda_2=\lambda_3=-1$ so that we have
\be
\sigma(\xi)=-\half\xi^3+(\ffrac{3}{2}-u)\xi^2+(\ffrac{3}{2}-u)\xi-\half.
\ee 
The first few terms of the Taylor series (\ref{eq:tay}) are
\be
C_0=0, \quad C_1=0, \quad C_2=0, \quad C_3=\ffrac{11}{6}+\ffrac{1}{6}u.
\ee
Once again, $C_3=0$ has no solution for $|u|<1$ so the method is
second-order.  The interval of periodicity ranges from $2^{-1/2}$ as
$u\to-1$ to 0 as $u\to1$.

For $k=4$, the distinct roots of $\xi$ can only be $\xi_1=1$,
$\xi_2=-1$ and $\xi_{3,4}=u\pm iv$ where $|u|<1$ and
$v=(1-u^2)^{1/2}$. Thus,
\be
\rho(\xi)=\xi^4-2u\xi^3+2u\xi-1.
\ee 
There are four choices for $\sigma(\xi)$: $\lambda_2\in\{-1,+1\}$,
$\lambda_3=\lambda_4\in\{-1,+1\}$. None of these yield methods of order $>2$ for
$|u|<1$.

The three- and four-step methods have no obvious advantage over the
trapezoidal or explicit midpoint methods, and we will not explore them
further.

\subsection{Five- and six-step methods}

\label{sec:fivesix}

For $k=5$, the distinct roots of $\xi$ are $\xi_1=1$,
$\xi_{2,3}=u_1\pm iv_1$ and $\xi_{4,5}=u_2\pm iv_2$, where $|u_j|<1$
and $v_j=(1-u_j^2)^{1/2}$. There are three choices for $\sigma(\xi)$:
$(\lambda_2,\lambda_4)=(-1,-1), (-1,+1)$, or $(+1,+1)$ (note that
$(+1,-1)$ is not a distinct choice), with $\lambda_3=\lambda_2$ and
$\lambda_5=\lambda_4$.  The choices $(\lambda_2,\lambda_4)=(-1,-1),
(+1,+1)$ yield only second-order methods. The choice
$(\lambda_2,\lambda_4)=(-1,+1)$ yields a fourth-order method if
\be u_2={1+11u_1\over 13-u_1}, \qquad u_1\in(-1,1), \qquad u_2\in
(-\ffrac{5}{7},1).
\label{eq:l5pm}
\ee
The integration formula is
\begin{eqnarray}
x_{n+1} & = & (x_n-x_{n-3})(1+2u_1+2u_2)-2(x_{n-1}-x_{n-2})(1+u_1+u_2+2u_1u_2)
+x_{n-4}\nonumber \\
& & \mbox{} +\ffrac{1}{2}h\big[f_{n+1}+(f_n+f_{n-3})(1+2u_1-6u_2)\nonumber \\
& & \mbox +2(f_{n-1}+f_{n-2})(1-3u_1+u_2+2u_1u_2)+f_{n-4}\big]
\qquad\qquad\hbox{method SZ5}.
\label{eq:s5pm}
\end{eqnarray}
The error constant is
\be
C={C_5\over\sigma(1)}={17u_1+103\over 1440(u_1-1)}.
\ee 
\begin{figure}
\vspace{12.5cm}
\includegraphics{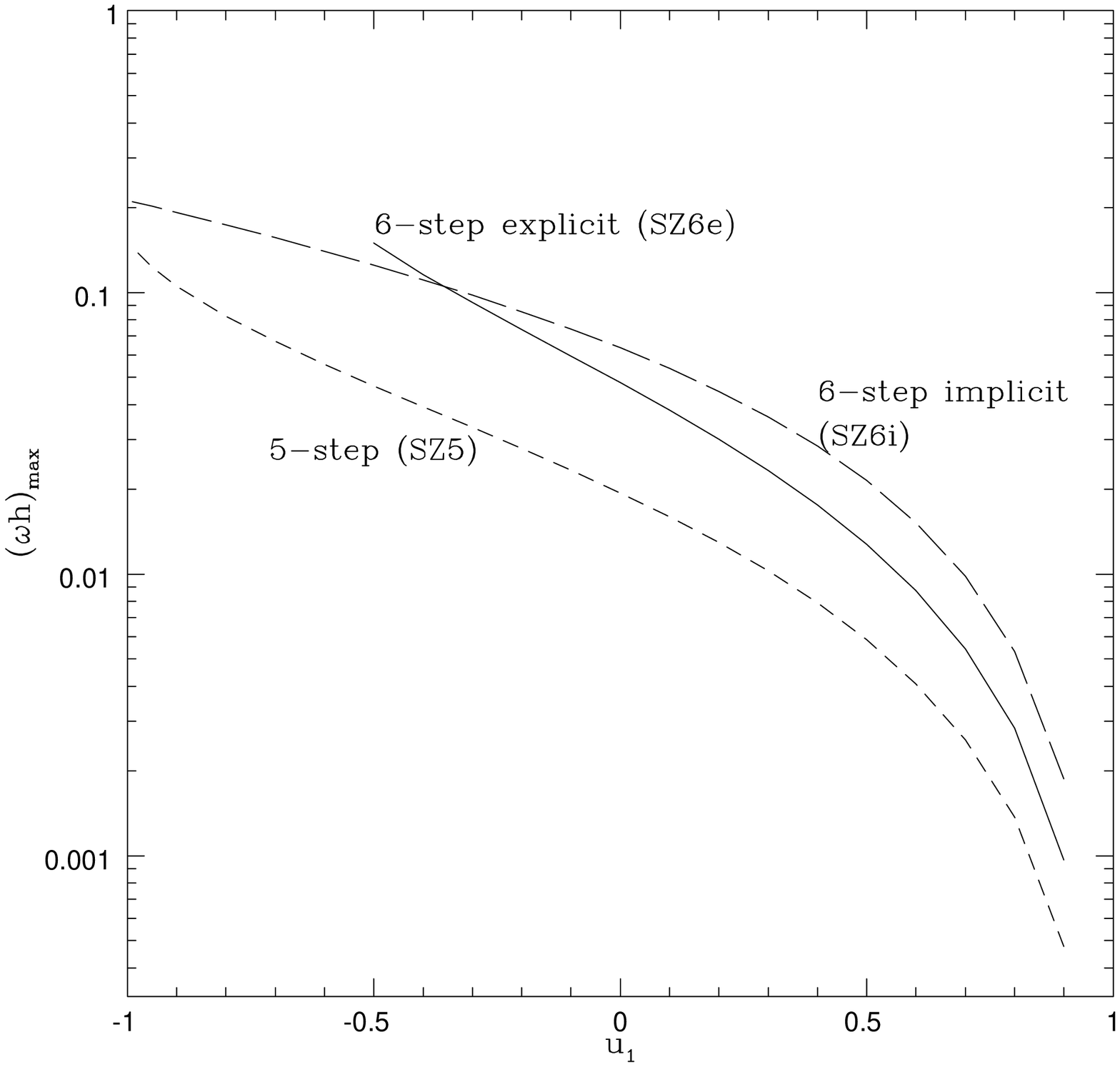}
\caption{The interval of periodicity for three zero-growth,
fourth-order \rias: the five-step method SZ5 defined by equations
(\ref{eq:l5pm}) and (\ref{eq:s5pm}), the 6-step implicit method SZ6i
defined by (\ref{eq:l6ppm}) and (\ref{eq:s6ppm}), and the 6-step
explicit method SZ6e defined by (\ref{eq:l6mpm}) and
(\ref{eq:s6mpm}).}
\label{fig:period}
\end{figure}

\begin{figure}
\vspace{12.5cm}
\includegraphics{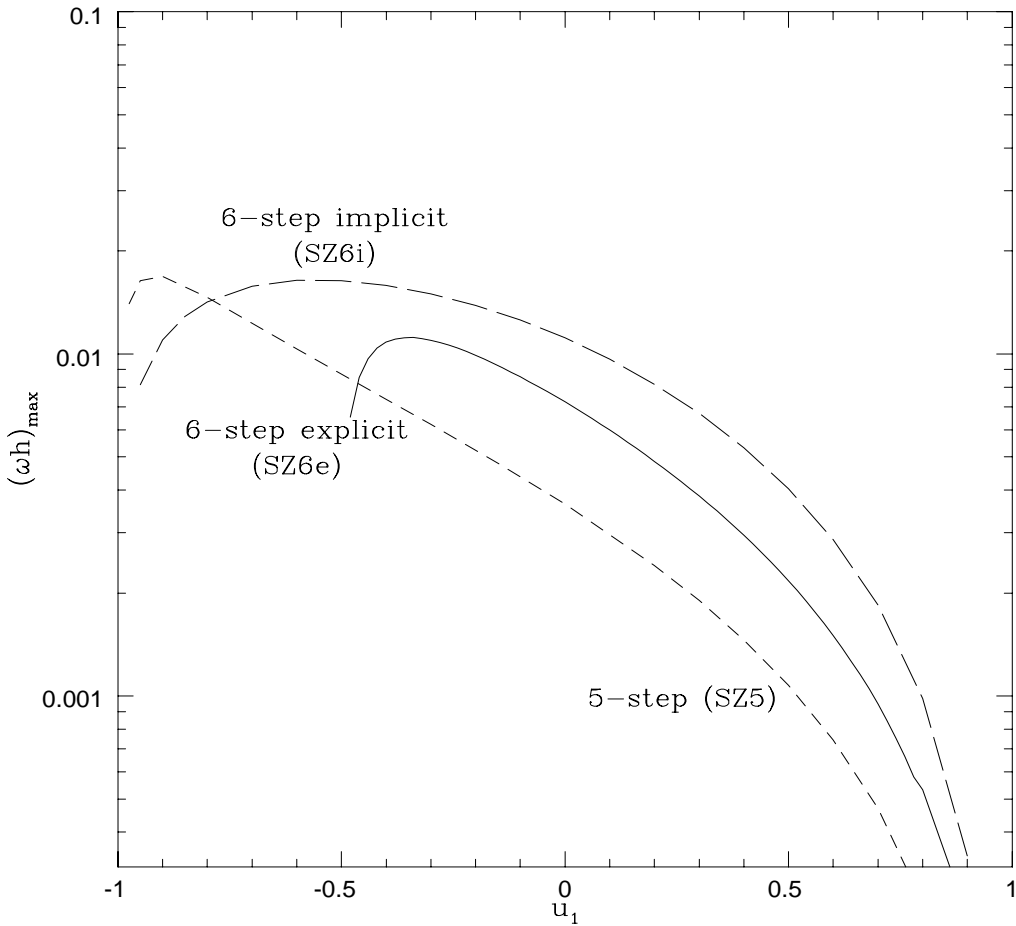}
\caption{The largest timestep that can integrate a circular orbit in
the reversible Kepler potential and maintain fractional energy error
below $10^{-7}$ for 100 orbital periods. Results are shown for the
three zero-growth, fourth-order \rias: the five-step method SZ5
defined by equations (\ref{eq:l5pm}) and (\ref{eq:s5pm}), the 6-step
implicit method SZ6i defined by (\ref{eq:l6ppm}) and (\ref{eq:s6ppm}),
and the 6-step explicit method SZ6e defined by (\ref{eq:l6mpm}) and
(\ref{eq:s6mpm}). Although results are displayed for circular orbits,
the curves for mildly eccentric orbits ($e \lta 0.1$) are almost
indistinguishable. This diagram can be compared with Figure
\ref{fig:period}, which shows the theoretical upper limit to the
timestep for the harmonic oscillator potential.}
\label{fig:periodkepler}
\end{figure}

For $k=6$, the distinct roots of $\xi$ are $\xi_1=1$, $\xi_2=-1$,
$\xi_{3,4}=u_1\pm iv_1$ and $\xi_{5,6}=u_2\pm iv_2$, where $|u_j|<1$
and $v_j=(1-u_j^2)^{1/2}$. There are six choices for $\sigma(\xi)$:
$\lambda_2=-1$ or $+1$ and $(\lambda_3,\lambda_5)=(-1,-1), (+1,-1)$,
$(+1,+1)$, with $\lambda_4=\lambda_3$ and $\lambda_6=\lambda_5$. The
choices $(\lambda_3,\lambda_5)=(-1,-1)$ or $(+1,+1)$ yield only
second-order methods.  The choice
$(\lambda_2,\lambda_3,\lambda_5)=(+1,+1,-1)$ yields a fourth-order
method if
\be
u_2={1+2u_1\over 4-u_1}, \qquad u_1\in(-1,1), \qquad u_2\in(-\ffrac{1}{5},1).
\label{eq:l6ppm}
\ee
The error constant is
\be
C={C_5\over\sigma(1)}={14+u_1\over 45(u_1-1)}. 
\ee 
The integration formula is
\begin{eqnarray}
x_{n+1} & = & 2(x_n-x_{n-4})(u_1+u_2)-(x_{n-1}-x_{n-3})(1+4u_1u_2)+x_{n-5}
\nonumber \\
& & \mbox{} +h\big[f_{n+1}+f_{n-5}-4(f_n+f_{n-4})u_2\nonumber \\
& & \mbox{   } +(f_{n-1}+f_{n-3})(3+4u_1u_2) -8f_{n-2}u_1\big] \qquad\qquad
\hbox{method SZ6i},
\label{eq:s6ppm}
\end{eqnarray}
the label ``i'' stands for ``implicit''. 

The choice $(\lambda_2,\lambda_3,\lambda_5)=(-1,+1,-1)$ yields an explicit fourth-order
method if
\be
u_2={7u_1-1\over u_1+5}, \qquad u_1\in(-\ffrac{1}{2},1), \qquad u_2\in(-1,1).
\label{eq:l6mpm}
\ee
The error constant is
\be
C={C_5\over\sigma(1)}={19+11u_1\over 180(1-u_1)}. 
\ee 
The integration formula is
\begin{eqnarray}
x_{n+1} & = & 2(x_n-x_{n-4})(u_1+u_2)-(x_{n-1}-x_{n-3})(1+4u_1u_2)+x_{n-5}
\nonumber \\
& & \mbox{}
+h\big[2(f_n+f_{n-4})(1+u_1-u_2)-4(f_{n-1}+f_{n-3})(u_1+u_2)+\nonumber \\
& & \mbox{   }4f_{n-2}(1-u_1+u_2+2u_1u_2)\big]\qquad\qquad \hbox{method SZ6e}.
\label{eq:s6mpm}
\end{eqnarray}
The intervals of periodicity for the three fourth-order methods are
shown in Figure \ref{fig:period}. These are largest for $u_1$ near
$-1$ and decrease rapidly for $u_1>0$. Thus, only methods with
negative $u_1$ are of practical use. At best, the interval of
periodicity is significantly smaller than that of comparable multistep
\rias\ for special second-order equations (Lambert \& Watson 1976;
Quinlan \& Tremaine 1990; Fukushima 1998). This is the price of the
greater generality of the present methods.

\section{Numerical experiments}

\label{sec:kep}

\begin{figure}
\vspace{12.5cm}
\includegraphics{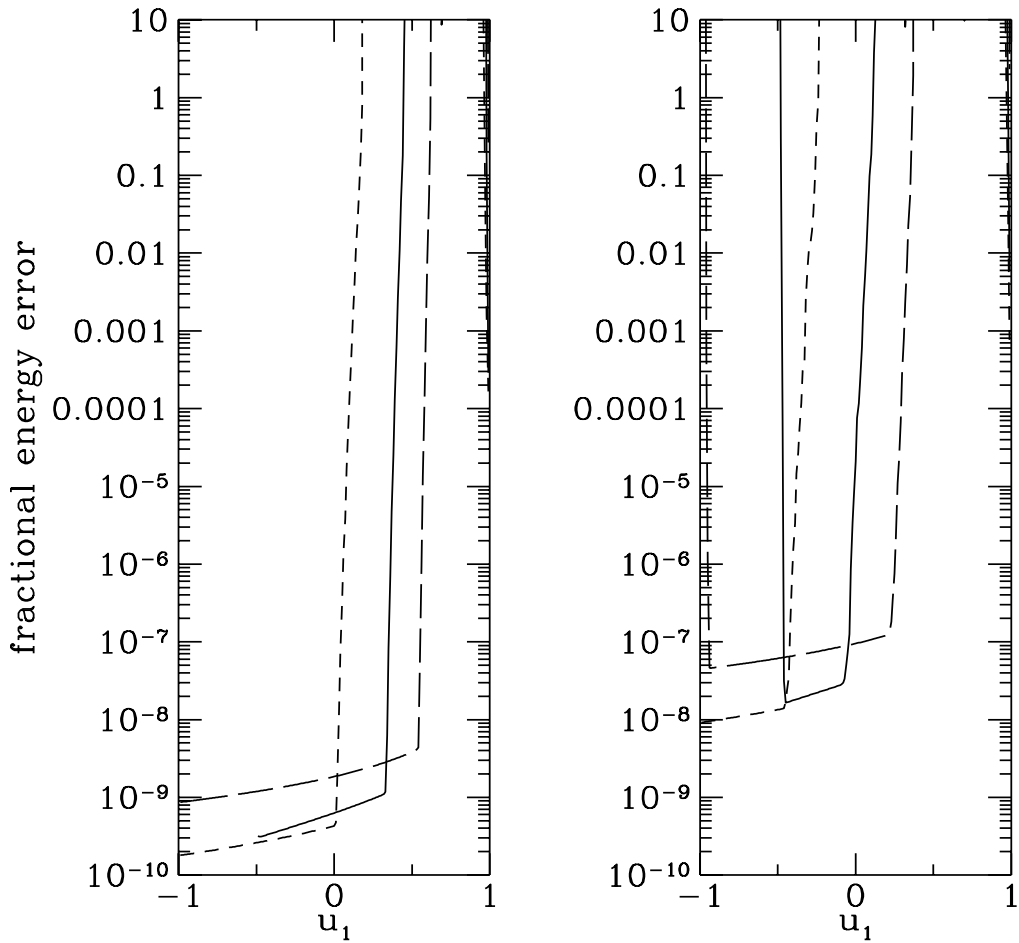}
\caption{The fractional energy error after integrating the standard
Kepler problem ($a=1$, $GM=1$) for 100 time units using timestep
$h=0.003$ and eccentricities 0.2 (left panel) and 0.5 (right
panel). The figure shows results for three fourth-order methods: the
five-step method SZ5 (short-dashed lines); the 6-step implicit method
SZ6i (long-dashed lines), and the 6-step explicit method SZ6e (solid
lines). The curves for $e=0.5$ are always higher than the
corresponding curves for $e=0.2$.}
\label{fig:multi56}
\end{figure}

This section supplements our theoretical analysis of multistep \rias\ with
numerical experiments using the Keplerian Hamiltonian ${1\over2}v^2-1/r$. The
equations of motion are
\be 
{dx\over dt}=v_x, \quad
{dy\over dt}=v_y, \quad {dv_x\over dt}=-{x\over r^3}, \quad {dv_y\over
dt}=-{y\over r^3}, 
\ee 
where $r^2=x^2+y^2$ and the gravitational constant $G$ and the attracting mass
$M$ have been set to unity. Our test integrations follow an orbit with
eccentricity $e$, unit semi-major axis, and orbital period $2\pi$. The orbit
is started at apocenter on the $x$-axis, so the initial conditions are
\be 
x=1+e, \quad y=0, \quad  v_x=0, \quad v_y=[(1-e)/(1+e)]^{1/2}.  
\ee 
The interval of periodicity shown in Figure \ref{fig:period} gives the 
maximum stable timestep for the harmonic oscillator potential only. It is of
greater relevance to astronomers to establish the timesteps that can
be safely used to integrate circular or mildly eccentric orbits in the
Kepler potential. Figure \ref{fig:periodkepler} shows the largest
timestep for which the relative energy error is less than $10^{-7}$
after 100 orbital periods. The results displayed are for circular
orbits, but the curves for mildly eccentric orbits ($e \lta 0.1$) are
almost indistinguishable. A comparison between Figures
\ref{fig:period} and \ref{fig:periodkepler} reveals that the timestep
needed for accurate integration of Keplerian near-circular orbits is
typically at least a factor of five smaller than would be naively
inferred from the interval of periodicity.

Figure \ref{fig:multi56} shows a further test of the multistep
\rias. The standard Kepler problem is integrated for 100 time units
with timestep $h=0.003$ and eccentricities 0.2 and 0.5. A striking
feature is that the error is almost independent of the parameter $u_1$
over a limited range, while outside this range the method is
unstable. The stable range shrinks as the eccentricity increases.
\begin{figure}
\vspace{12.5cm}
\includegraphics{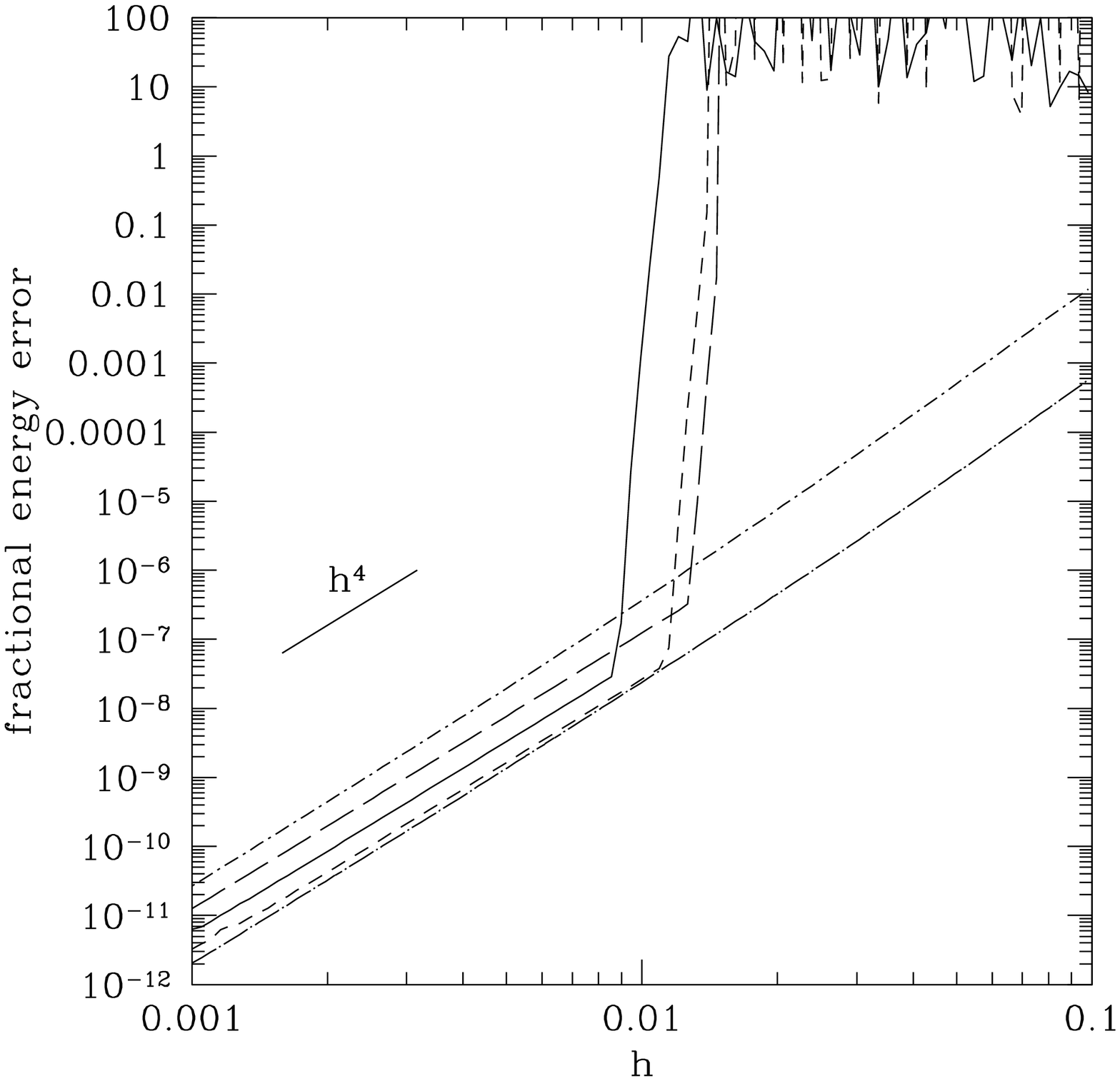}
\caption{The fractional energy error after integrating the standard
Kepler problem ($a=1$, $GM=1$) with eccentricity $e=0.2$ for 100 time
units.  The figure shows results for three zero-growth, fourth-order
\rias: the five-step method SZ5 with $u_1=-0.75$ (short-dashed line);
the 6-step implicit method SZ6i with $u_1=-0.75$ (long-dashed line);
and the 6-step explicit method SZ6e with $u_1=-0.25$ (solid
line). Errors are also shown for two other fourth-order multistep
methods: Adams-Bashforth (dot-short dashed line) and Adams-Moulton
(dot-long dashed line).}
\label{fig:hvar56}
\end{figure}
The dependence of energy error on timestep is explored in Figure
\ref{fig:hvar56}. In this Figure, we have specialized to a single
value of the parameter $u_1$: $-0.75$ for SZ5 and SZ6i and $-0.25$ for
SZ6e. All three methods display the expected dependence $\Delta
E\propto h^4$ for $h<0.01$. For larger values of $h$, the methods are
unstable. We have also plotted the behavior of the classic
fourth-order Adams-Bashforth (explicit) and Adams-Moulton (implicit)
multistep methods. The classic methods exhibit similar errors for
$h<0.01$ but remain stable at much larger timesteps.
\begin{figure}
\vspace{12.5cm}
\includegraphics{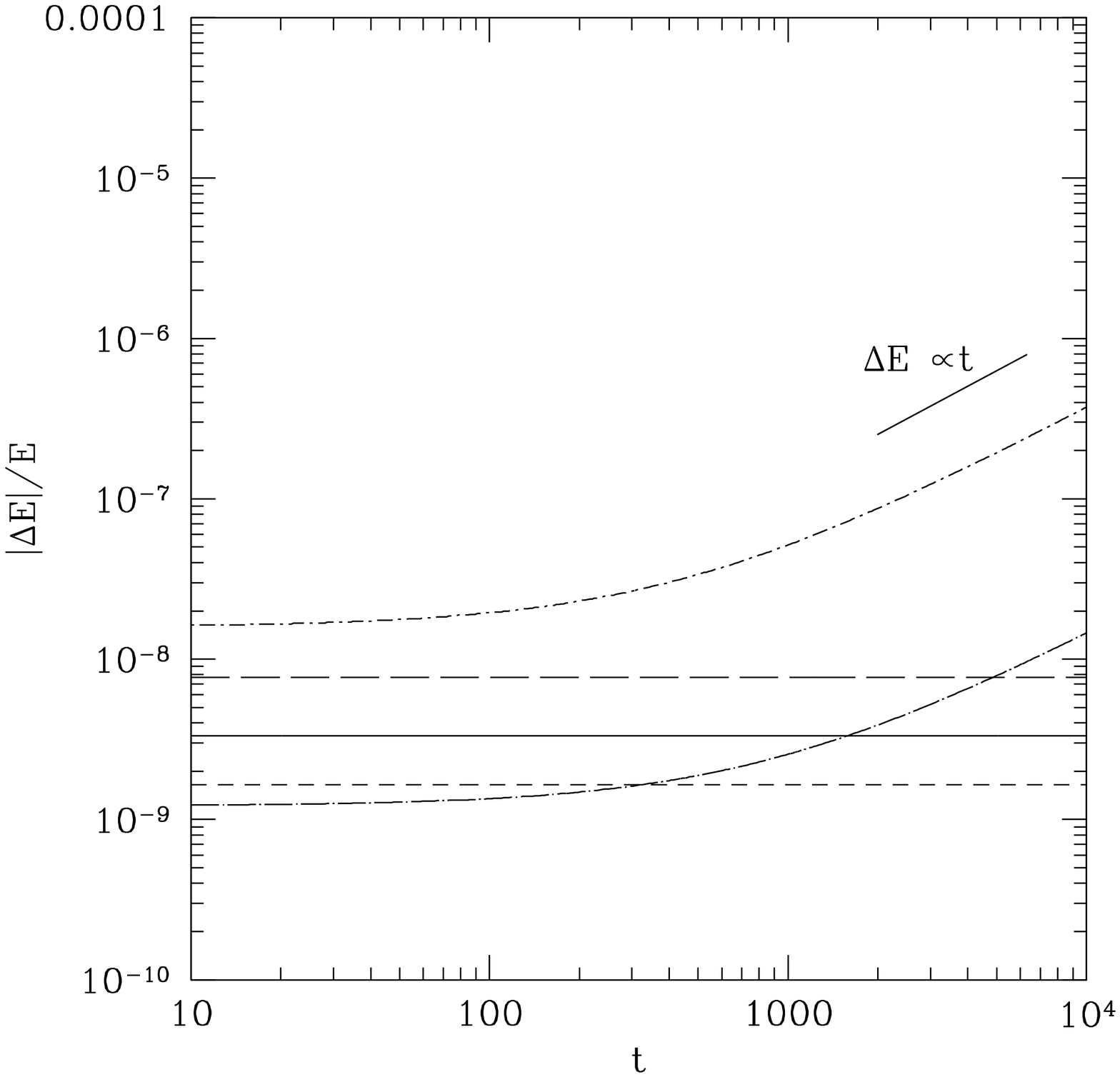}
\caption{The maximum fractional energy error over integration time $t$
for the standard Kepler problem ($a=1$, $GM=1$) with eccentricity
$e=0.2$ and timestep $h=0.005$.  The figure shows results for three
zero-growth, fourth-order \rias: the five-step method SZ5 with
$u_1=-0.75$ (short-dashed line); the 6-step implicit method SZ6i with
$u_1=-0.75$ (long-dashed line); and the 6-step explicit method SZ6e
with $u_1=-0.25$ (solid line). Errors are also shown for the
fourth-order Adams-Bashforth (dot-short dashed line) and Adams-Moulton
methods (dot-long dashed line).}
\label{fig:long56}
\end{figure}
The advantage of the reversible multistep methods only becomes clear over
longer time intervals. Figure \ref{fig:long56} shows that the maximum
energy error in the reversible methods is constant at large times,
while the energy error in the classical methods grows linearly.

\subsection{Variable timesteps}

\begin{figure}
\vspace{12.5cm}
\includegraphics{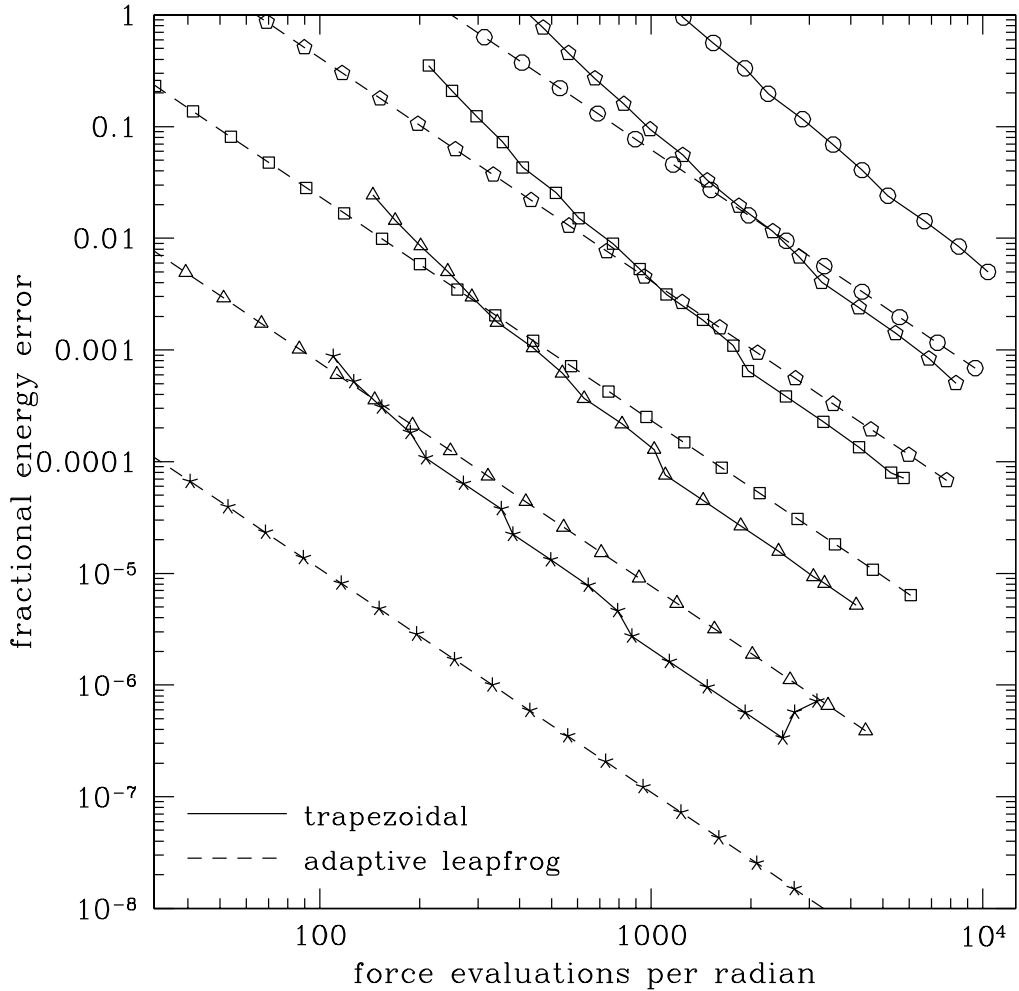}
\caption{Results of integrating the differential equations for a Kepler orbit
(\ref{eq:kepvar}) for 1000 time units, using the trapezoidal method
(eq. \ref{eq:trap}, solid lines) and explicit adaptive leapfrog
(eq. \ref{eq:calvo}, dashed lines). The orbit eccentricities are $e=0.5$
(stars), 0.9 (triangles), 0.99 (squares), 0.999 (pentagons), 0.9999 (circles).
The vertical axis is the maximum fractional energy error during the
integration and the horizontal axis is the number of evaluations of the
right-hand side of equations (\ref{eq:kepvar}) per unit time (the orbital
period is $2\pi$). }
\label{fig:trap}
\end{figure}

Variable timesteps are useful for eccentric Kepler orbits.  To
incorporate these, we introduce a fictitious time $\tau$
by the relation $dt=g(x,y)d\tau$, so the equations of motion become
\be 
{dx\over d\tau}=g(x,y)v_x, \quad {dy\over d\tau}=g(x,y)v_y,
\quad {dv_x\over d\tau}=-g(x,y){x\over r^3}, \quad {dv_y\over
d\tau}=-g(x,y){y\over r^3}, \quad {dt\over d\tau}=g(x,y).
\label{eq:kepvar}
\ee 
These are integrated using unit timestep in $\tau$. We typically use
$g(x,y)\propto r^{3/2}$ since this is the characteristic free-fall time from
radius $r$. We parameterize the timestep by the number of force evaluations
per unit time (or per radian, since the orbital period is $2\pi$). This is
distinct from the number of steps per radian because implicit methods require
several iterations per step, and because some explicit methods -- though not
multistep methods -- require several force evaluations per step.  Figure
\ref{fig:trap} shows the fractional energy errors resulting from integrating
Kepler orbits using the trapezoidal method (eq. \ref{eq:trap}, solid lines)
and explicit adaptive leapfrog (eq. \ref{eq:calvo}, dashed lines). Each step
of the trapezoidal method was iterated to convergence, with the first
approximation taken from the first-order Euler method -- typically 3
iterations were required at the smallest timesteps, rising to $\sim 10$ at the
largest.  The energy errors are the maximum over 1000 time units or $500/\pi$
orbits. The results for shorter integrations over 100 time units are almost
indistinguishable, showing that there is no secular energy drift with either
integration method. Adaptive leapfrog generally gives energy errors that are
smaller by about an order of magnitude, mostly because it is explicit so there
are fewer force evaluations per timestep.  We have conducted similar
experiments with the explicit midpoint method (eq. \ref{eq:mid}, not shown in
Figure) but this is much less successful at following high-eccentricity
orbits, presumably because of its smaller interval of periodicity.

\begin{figure}
\vspace{12.5cm}
\includegraphics{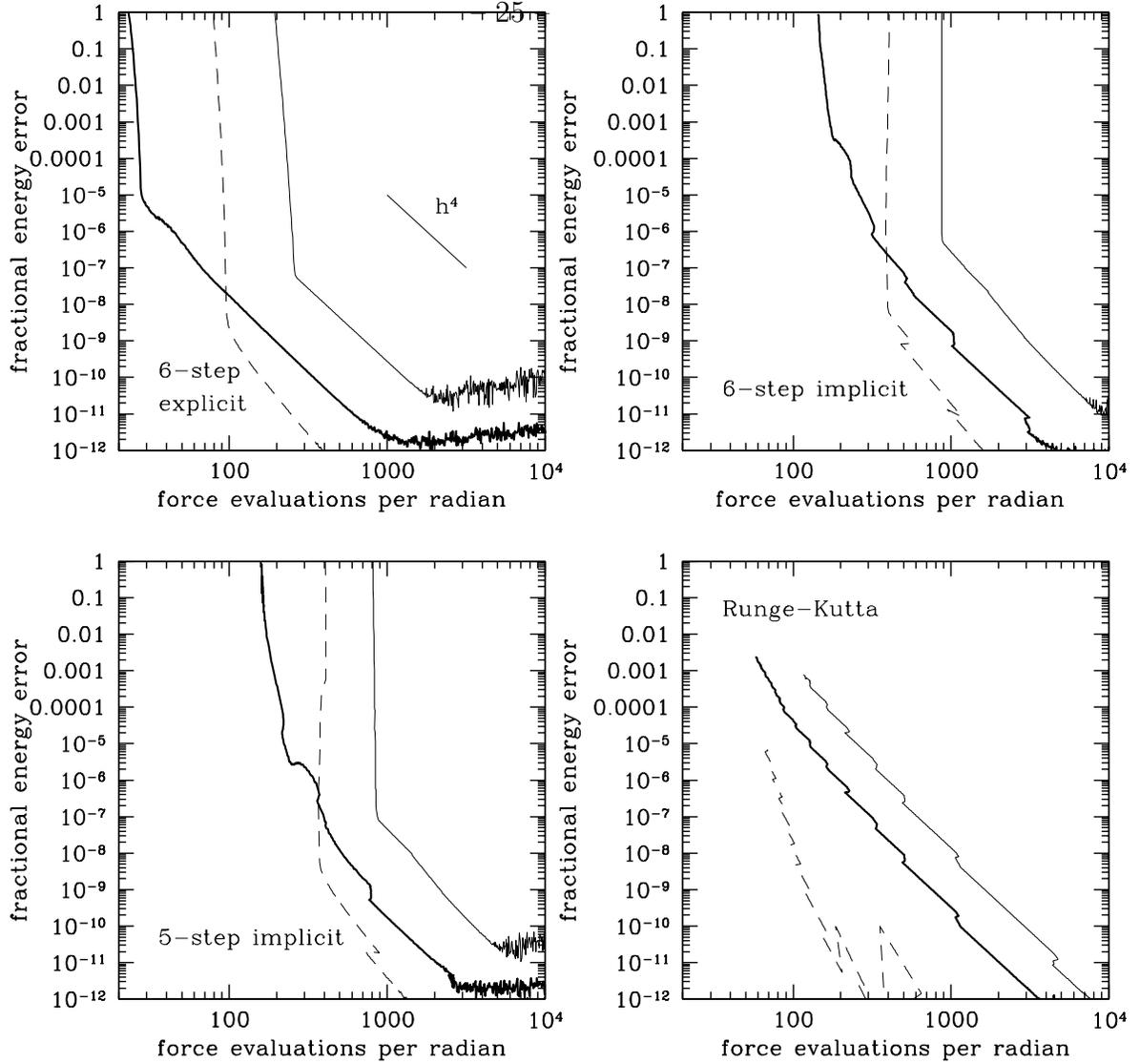}
\caption{Integration of a Kepler orbit with eccentricity $e=0.5$ by
four fourth-order methods: the explicit six-step method with
$u_1=-0.25$ (method SZ6e, eq. \ref{eq:s6mpm}); the implicit six-step
method with $u_1=-0.75$ (method SZ6i, eq. \ref{eq:s6ppm}); the
implicit five-step method with $u_1=-0.75$ (method SZ5,
eq. \ref{eq:s5pm}); and the implicit, fourth-order, two-stage
Runge-Kutta \ria\ defined by equations (\ref{eq:rk}).  The heavy solid
lines are the fractional energy errors for $e=0.5$ and variable
timestep, the light solid lines are for $e=0.5$ and fixed timestep,
and the dashed lines are for $e=0$ and fixed timestep.}
\label{fig:fourth}
\end{figure}

We next investigate the fourth-order multistep \rias. These have
relatively small intervals of periodicity and cannot reliably
integrate high-eccentricity orbits. However, they are successful at
following orbits with moderate eccentricities.  Figure
\ref{fig:fourth} shows the fractional energy errors resulting from
integrating a Kepler orbit with eccentricity $e=0.5$ using four
methods: the explicit six-step method with $u_1=-0.25$ (method SZ6e,
eq. \ref{eq:s6mpm}); the implicit six-step method with $u_1=-0.75$
(method SZ6i, eq. \ref{eq:s6ppm}); the implicit five-step method with
$u_1=-0.75$ (method SZ5, eq. \ref{eq:s5pm}).  In all panels, the heavy
solid line is the error resulting from integrating the eccentric orbit
with variable timestep, the light solid line is the error from
integrating the same orbit with fixed timestep, and the dashed line is
the error from integrating a circular orbit with fixed timestep. For
large timesteps, the implicit methods do not converge and in this case
no energy error is plotted.
For all four methods a variable timestep yields errors that are 1--2
orders of magnitude smaller than a fixed timestep (at the same number
of function evaluations per unit time). The explicit multistep method
performs much better than the implicit methods, in part because the
implicit methods require 3--8 force evaluations per step to iterate to
convergence. However, even if the convergence of the implicit
multistep methods could be achieved in 2--3 iterations they would not
perform significantly better than the explicit method on this test.  A
surprising result is that there is a range of the horizontal axis in
which the multistep methods with variable timestep integrate eccentric
orbits with smaller energy error than circular orbits.

\begin{figure}
\vspace{12.5cm}
\includegraphics{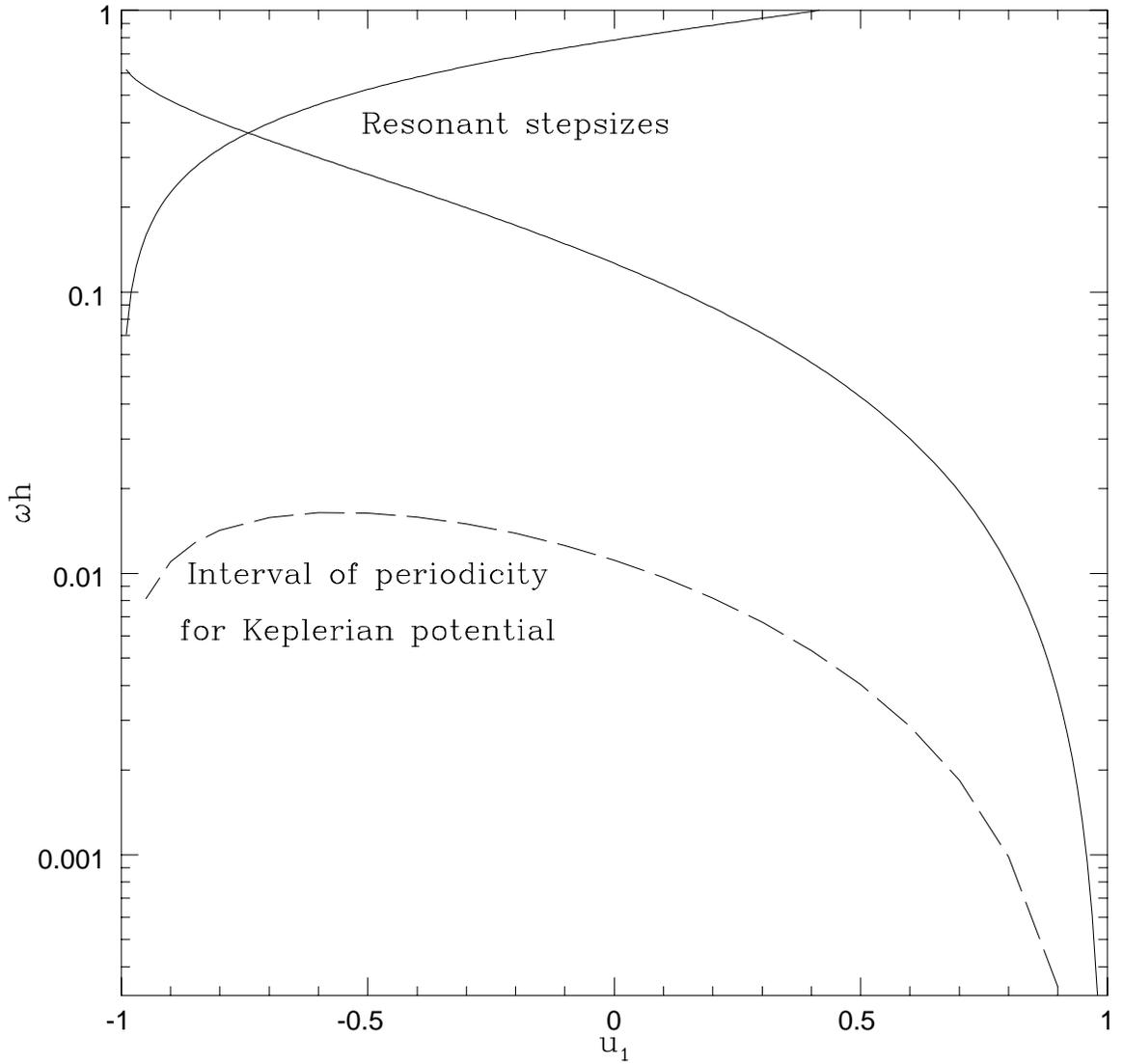}
\caption{For the 6-step implicit method SZ6i, the resonant
timesteps are shown in the full lines. The interval of periodicity
for nearly circular Keplerian orbits is shown in a dashed line.
The resonant timesteps are always larger than the maximum permitted
timestep. Similar results hold true for SZ5 and SZ6e.
}
\label{fig:final}
\end{figure}

\subsection{Resonant instabilities}

Toomre (1990, private communication) and Quinlan (1999) showed that
multistep \rias\ for the special second-order differential equation
(\ref{eq:fff}) suffer instabilities at unlucky timesteps at which
there is a resonance between the oscillation frequencies of the
solution and the method. This phenomenon must also occur in the
multistep \rias\ for first-order differential equations that are
investigated here.  Suppose the angular speed of the circular orbit
given by the numerical method is $\omega$. If the number of steps in
the multistep \ria\ is at least 5 (or 6 for second-order differential
equations), then there are two or more complex-conjugate pairs of
roots of the characteristic polynomial (\ref{eq:poly}) on the unit
circle. These we write as $\xi_i = \exp(\pm i\theta_i)$ and $\xi_j =
\exp(\pm i\theta_j)$. Quinlan (1999) shows that the troublesome
timesteps $h$ satisfy
\be
\theta_i - \theta_j = 2\omega h.
\label{eq:gerry}
\ee

In practice these resonant instabilities are never a concern for our
zero-growth multistep \rias. This is because the timesteps
(\ref{eq:gerry}) lie outside the interval of periodicity for nearly
circular Keplerian orbits.  Let us demonstrate this by considering one
example in more detail -- namely, the 6-step implicit method SZ6i
defined by (\ref{eq:l6ppm}) and (\ref{eq:s6ppm}). This is the most
susceptible to resonant instabilities of the zero-growth methods we
have examined, both because it has the largest interval of periodicity
and because, as $u_1 \rightarrow -1$, two of the spurious roots
coalesce as $u_1\rightarrow -1$.  We recall that there are five
spurious roots on the unit circle, namely $\xi_2=-1$, and the two
complex-conjugate pairs $\xi_{3,4}=u_1\pm iv_1$ and $\xi_{5,6}=u_2\pm
iv_2$, where $v_j=(1-u_j^2)^{1/2}$.  Using (\ref{eq:l6ppm}), the
angular separations between the spurious roots on the unit circle can
be readily computed as a function of $u_1$. Figure \ref{fig:final}
shows the locations of the potentially troublesome timesteps $h$,
together with the maximum possible timestep (already presented in
Figure \ref{fig:periodkepler}). The resonant instabilities always
occur at timesteps that are outside the interval of periodicity for
Keplerian orbits. Similar results hold for the methods SZ5 and SZ6e.

\section{Conclusions}

\label{sec:disc}

The generic instabilities of multistep \rias\ for first-order differential
equations identified by Cano \& Sanz-Serna (1998) are a grave problem and
render many of these methods unusable. The main contribution of this paper is
the elaboration of a class of multistep \rias\ -- the zero-growth methods --
that successfully evade the instabilities. This select group includes two
familiar second-order multisteps, namely the trapezoidal method
(\ref{eq:trap}) and the explicit midpoint method (\ref{eq:mid}), as well as
entirely new multistep \rias. In particular, we have identified three
one-parameter families of zero-growth, fourth-order \rias\ -- namely, the
five-step implicit method SZ5 defined by equations (\ref{eq:l5pm}) and
(\ref{eq:s5pm}), the six-step implicit method SZ6i defined by (\ref{eq:l6ppm})
and (\ref{eq:s6ppm}), and the six-step explicit method SZ6e defined by
(\ref{eq:l6mpm}) and (\ref{eq:s6mpm}). Of these, the explicit method SZ6e
eliminates the need to iterate to convergence at each timestep without a major
sacrifice in the interval of periodicity. The stability of the zero-growth
methods has been understood at a theoretical level and confirmed at a
practical level by integrations of circular and eccentric Keplerian orbits. 

There are several advantages of multistep \rias\ for first-order
differential equations over the multistep \rias\ for special
second-order differential equations examined by Lambert \& Watson
(1976), Quinlan \& Tremaine (1990), and Fukushima (1998, 1999). They
permit the introduction of variable timesteps without spoiling
reversibility or long-term energy conservation. They are also more
general, since any ordinary differential equations can always be
reduced to a system of first-order equations; thus, for example, the
methods described here can be used to follow orbits in rotating frames
of reference or the motions of rotating rigid bodies. A third
advantage is that they are less susceptible to timestep resonances.
The price paid for these attractive features is that (i) the interval
of periodicity of the higher order methods such as SZ5, SZ6i and SZ6e
is rather small, forcing the use of smaller timesteps than other
methods; (ii) more steps are needed for a method of given order; there
are no fourth-order methods with fewer than five steps and no fifth or
sixth-order methods with 8 or fewer steps (the limit of our
explorations).  Nonetheless, the six-step explicit method (SZ6e)
remains a competitive option for problems in which it is critical to
have variable timesteps and to avoid irreversible numerical errors:
examples include long-term integrations of asteroid or comet orbits,
or orbits near the centers of galaxies.

There are several possible avenues for future exploration:

\begin{enumerate}

\item Are there zero-growth \rias\ with more than $k=6$ steps that have
higher order or a larger interval of periodicity? We have explored all $k\le
8$ but have found no \rias\ of order greater than 4.

\item Gautschi (1961) developed a class of methods closely
related to the linear multisteps.  Just as a linear multistep is
defined by the requirement that the operator (\ref{eq:linearoperator})
annihilate all algebraic polynomials of order $\le p$, so the Gautschi
multisteps annihilate certain trigonometric polynomials up to a given
degree.  These methods are particularly appealing when the solution is
known to be periodic and a reasonable estimate for the period of the
orbit can be guessed in advance -- requirements that are often
satisfied in solar system integrations. Can we find reversible
Gautschi multisteps?

\item Integration methods are generally designed to maximize
the order $p$, defined so that the one-step error in following a
system with characteristic frequency $\omega$ is proportional to
$(h\omega)^{p+1}$. Perhaps it is more sensible to
``economize'' the error, by minimizing the maximum value of the error
over a range of frequencies $0<\omega\le\omega_{\rm max}$. In this
case the error would be nearly proportional to a Chebyshev polynomial
$T_{p+1}(x)$ with argument $x=\omega/\omega_{\rm max}$ and degree
$p+1$, instead of $x^{p+1}$.

\end{enumerate}

\begin{acknowledgments}

We are indebted to Gerry Quinlan for communicating his results on
resonant instabilities in advance of publication.  This research was
supported in part by NASA grant NAG5-7310. NWE thanks both the
Canadian Institute for Theoretical Astrophysics and the Department of
Astrophysical Sciences, Princeton University for their hospitality
during working visits.  His research is supported by the Royal
Society.

\end{acknowledgments}

\end{document}